\documentstyle[twoside,fleqn,espcrc2,epsfig]{article}

\newcommand{\AmS}{{\protect\the\textfont2
  A\kern-.1667em\lower.5ex\hbox{M}\kern-.125emS}}

\title{Effects of Quenching and Partial Quenching on QCD 
Penguin Matrix Elements} 

\author{ Maarten Golterman\address{Dept. of Physics and Astronomy, 
San Francisco State University, San Francisco, CA 94132, USA 
 }
and 
Elisabetta Pallante\address{S.I.S.S.A., Via Beirut 2-4, 34014 Trieste, Italy}%
\thanks{Talk given by E. Pallante at Lattice 2001} } 
       
\begin{document}

\newcommand{\be}{\begin{equation}}
\newcommand{\ee}{\end{equation}}
\newcommand{\ba}{\begin{eqnarray}}
\newcommand{\ea}{\end{eqnarray}}

\newcommand{\cL}{{\cal L}}
\newcommand{\cM}{{\cal M}}
\newcommand{\Bt}{{\tilde B}}
\newcommand{\cO}{{\cal O}}
\newcommand{\cOt}{{\tilde\cO}}
\newcommand{\bt}{{\tilde\beta}}
\newcommand{\tr}{{\mbox{tr}\,}}
\newcommand{\str}{{\mbox{str}\,}}
\newcommand{\Exp}{{\mbox{exp}\,}}
\newcommand{\Mdot}{{\dot M}}
\newcommand{\Mbar}{{M_{VS}}}
\newcommand{\tb}{{\tilde\beta}}
\newcommand{\vp}{{\vec p}}
\newcommand{\hX}{{\hat X}}
\newcommand{\diag}{{\rm diag}}
\newcommand{\sbar}{{\overline{s}}}
\newcommand{\dbar}{{\overline{d}}}
\newcommand{\ubar}{{\overline{u}}}
\newcommand{\qbar}{{\overline{q}}}
\newcommand{\psibar}{{\overline{\psi}}}
\newcommand{\ie}{{\it i.e.}}
\newcommand{\Nh}{{\hat N}}

\begin{abstract}

We point out that chiral transformation properties of penguin operators 
change in the transition from unquenched to (partially) quenched QCD.
The way in which this affects 
the lattice determination of weak matrix elements can be understood 
in the framework of (partially) quenched chiral perturbation theory.
\end{abstract}

\maketitle

\section{(Partially) Quenched QCD penguins}

We consider $\Delta S=1$ $LR$ operators of the form
\be
Q^{QCD}_{penguin}=(\sbar d)_L(\qbar q)_R\ , \label{penguin}
\ee
with $q=u,d,s$,
$(\qbar_1 q_2)_{L,R}=\qbar_1\gamma_\mu P_{L,R}q_2$,
and $P_{L,R} = (1\mp \gamma_5)/2$ left- and right-handed projectors.
Color indices can be contracted in two ways, 
corresponding to the QCD penguins $Q_{5,6}$. 

The operators in eq.~(\ref{penguin}) are obtained by the unquenched QCD 
evolution of the weak operator from the weak scale $\sim M_W$ down to the 
hadronic scale $\sim m_c$. At hadronic scales, 
matrix elements of these four-quark operators are usually 
computed on the lattice in quenched or partially quenched QCD
((P)QQCD). The key point we wish to address here is that the chiral 
transformation properties 
of QCD penguin operators change in the transition from unquenched to 
(P)QQCD. This has non-trivial consequences \cite{OUR}
which can be understood in the 
framework of (partially) quenched chiral perturbation theory ((P)QChPT) 
\cite{bgpq}.

PQQCD can be systematically formulated in a lagrangian
framework by coupling the 
gluons to three sets of quarks \cite{bgpq}: $K$
valence quarks $q_{vi}$ with masses $m_{v1},m_{v2},\dots,m_{vK}$, $N$
sea quarks $q_{si}$ with masses $m_{s1},m_{s2},\dots,m_{sN}$, and $K$
ghost quarks $q_{gi}$ with masses $m_{v1},m_{v2},\dots,m_{vK}$.
The ghost quarks are identical to the valence quarks, except for their
statistics, which is chosen to be bosonic \cite{morel}.
Quenched QCD corresponds to the
special case $N=0$ (no sea quarks).  
It can be shown that unquenched QCD below the charm threshold 
corresponds to the choice $K=N=3$ and $m_{si}=m_{vi}$, $i=1,\dots,3$
\cite{bgpq,shsh}. 

The total number of quarks in PQQCD is thus $2K+N$, and correspondingly,
the chiral symmetry group enlarges from $SU(3)_L\times SU(3)_R$
to the graded group $SU(K+N|K)_L\times SU(K+N|K)_R$.
As a consequence, the operators of eq.~(\ref{penguin}), which
are singlets under $SU(3)_R$, are no longer singlets
under the enlarged group $SU(K+N|K)_R$.
They can be decomposed as 
\ba
Q^{QCD}_{penguin}&=&
\frac{K}{N}\;\str(\Lambda\psi\psibar\gamma_\mu P_L)
\;\str(\psi\psibar\gamma_\mu P_R)+  \nonumber \\
&&\str(\Lambda\psi\psibar\gamma_\mu P_L)
\;\str(A\psi\psibar\gamma_\mu P_R)\ , \nonumber \\
&\equiv&\frac{K}{N}\;Q^{PQS}_{penguin}+Q^{PQA}_{penguin}\ , \label{pqdecomp}
\ea
\be
A=\diag(1-\frac{K}{N},\dots,1-\frac{K}{N},
-\frac{K}{N},\dots,-\frac{K}{N})\ , \label{a}
\ee
where the first $K$ (valence) entries of $A$ are equal to $1-{K}/{N}$,
and the next $N+K$ (sea and ghost) entries are equal to $-{K}/{N}$.
The superscripts $PQS$ and $PQA$ indicate that these operators transform
in the singlet and adjoint representations of $SU(K+N|K)_R$, respectively.
In the quenched case the situation is special.
The decomposition reads
\ba
Q^{QCD}_{penguin}&=&
\frac{1}{2}\;\str(\Lambda\psi\psibar\gamma_\mu P_L)
\;\str(\psi\psibar\gamma_\mu P_R)+\nonumber \\
&&\str(\Lambda\psi\psibar\gamma_\mu P_L)
\;\str(\Nh\psi\psibar\gamma_\mu P_R)\ , \nonumber \\
&\equiv&\frac{1}{2}\;Q^{QS}_{penguin}+Q^{QNS}_{penguin}\ , \label{qdecomp}\\
\Nh&=&\frac{1}{2}\diag(1,\dots,1,
-1,\dots,-1)\ , \label{s}
\ea
where the first $K$ (valence) entries of $\Nh$ are equal to $\frac{1}{2}$,
and the last $K$ (ghost) entries are equal to $-\frac{1}{2}$.  The first
operator in the decomposition is a singlet, while the second is not,
under $SU(K|K)_R$ ($NS$ for non-singlet).  
However, the unit matrix  has now a vanishing
supertrace, while $\Nh$ has not, and $Q^{QNS}_{penguin}$ can mix with 
$Q^{QS}_{penguin}$ through penguin-like diagrams.

\section{Representation of QCD penguins in PQChPT  }

QCD penguins transform as $(8,1)$ under the chiral symmetry group
and start at order $p^2$ in ordinary ChPT 
(see {\it e.g.} ref.~\cite{dghcbtasi}).
Denoting the adjoint representation of the PQ group by $A$, we found
in the previous section that $Q^{PQS}_{penguin}$ transforms as
$(A,1)$, while $Q^{PQA}_{penguin}$ transforms as
$(A,A)$ under $SU(K+N|K)_L\times SU(K+N|K)_R$.  
To lowest order in PQChPT and in euclidean space, 
these operators are represented by
\ba
Q^{PQS}_{penguin}&\hspace{-0.3cm}\rightarrow\hspace{-0.3cm}& 
-\alpha^{(8,1)}_1\str(\Lambda L_\mu L_\mu)
+\alpha^{(8,1)}_2\str(\Lambda X_+) \nonumber \\
Q^{PQA}_{penguin}&\hspace{-0.3cm}\rightarrow\hspace{-0.3cm}&
f^2\;\alpha^{(8,8)}\;
\str(\Lambda\Sigma A\Sigma^\dagger)
\ , \label{pqa} 
\ea
where $L_\mu=i\Sigma\partial_\mu\Sigma^\dagger$,  
$X_\pm=2B_0(\Sigma M^\dagger\pm M\Sigma^\dagger)$, 
with $M$ the quark-mass matrix, $B_0$ the parameter $B_0$ of ref.~\cite{gl},
$\Sigma=\Exp(2i\Phi/f)$ describing the partially-quenched
Goldstone-meson multiplet, and $f$ the bare pion-decay constant
normalized such that $f_\pi=132$~MeV.  The low-energy constants (LECs)
$\alpha^{(8,1)}_{1,2}$ also appear 
in the unquenched theory. However, the LEC
$\alpha^{(8,8)}$ only appears in the PQ case, and multiplies an
order $p^0$ operator, which is possible because $A$ is non-trivial.  
It has a direct physical meaning, because the same LEC also appears
in the bosonization of the EM penguin (which belongs to 
the same irreducible representation as $Q^{PQA}_{penguin}$).
Quenched bosonization rules to leading order are given by
\ba
Q^{QS}_{penguin}&\hspace{-0.3cm}\rightarrow\hspace{-0.3cm}&
 -\alpha^{(8,1)}_{q1}\str(\Lambda L_\mu L_\mu)
+\alpha^{(8,1)}_{q2}\str(\Lambda X_+) \nonumber \\
Q^{QNS}_{penguin}&\hspace{-0.3cm}\rightarrow\hspace{-0.3cm}& 
f^2\;\alpha^{NS}_q\;
\str(\Lambda\Sigma \Nh\Sigma^\dagger) \; .  \label{qchpt}
\ea
The quenched case differs from the partially quenched one in two ways.  
First, all ChPT LECs depend on $N$, and thus their quenched values are 
not necessarily equal to their $N\ne 0$ values.
Second, in this case the operator $Q^{QNS}_{penguin}$ 
and the EM penguin do not belong to the same irrep and their LECs
are in principle not related \cite{OUR}. 

\section{Kaon matrix elements in PQChPT}

Since the new operators $Q^{PQA}_{penguin}$ in eq.~(\ref{pqa}) and 
 $Q^{QNS}_{penguin}$  in eq.~(\ref{qchpt})
are of order $p^0$ in ChPT (while singlet LR operators are of order $p^2$), 
they potentially lead to an enhancement relative to the unquenched case.
As the simplest example, we consider $K\to\pi$ and 
$K\to 0$ matrix elements, to leading order in ChPT.  
New contributions also show up
in direct $K\to\pi\pi$ matrix elements \cite{gpll}. 

It turns out that $Q^{PQA}_{penguin}$ (and $Q^{QNS}_{penguin}$)
do not contribute to any of these matrix elements (i.e. $K\to\pi$, 
$K\to 0$ and $K\to\pi\pi$)  at order $p^0$,  
but they do in general contribute at order $p^2$.  
Since  $Q^{PQS}_{penguin}$ starts at order $p^2$, the new
contributions from $Q^{PQA}_{penguin}$
compete at the {\it leading} order of the chiral expansion
of these matrix elements, 
and have to be taken into account even if one analyzes lattice results 
using only leading-order ChPT.  
For the PQ $K\to\pi$ matrix element, with 
degenerate valence quark masses ($M^2=M_K^2=M_\pi^2=2B_0 m_v$),
we find at order $p^2$
\ba
&\hspace{-0.5cm}&[K^+\to\pi^+]^{QCD}_{penguin}
=\frac{4M^2}{f^2}\Biggl\{\alpha^{(8,1)}_1-\alpha^{(8,1)}_2 - 
\nonumber\\
 &\hspace{-0.5cm}&\frac{2}{(4\pi)^2}\left(1-\frac{K}{N}\right)
(\beta^{(8,8)}_1+\frac{1}{2}\beta^{(8,8)}_2+\beta^{(8,8)}_3 )
\Biggr\} \ , \label{ktopi} 
\ea
where the LECs $\beta^{(8,8)}_{1,2,3}$ appear in the bosonization of 
$Q^{PQA}_{penguin}$ at order $p^2$ \cite{OUR}.
The $K\to 0$ matrix element with arbitrary valence
and sea quarks is
\ba
&&\hspace{-0.6cm}[K^0\to 0]^{QCD}_{penguin}
=\frac{4i}{f}\Biggl\{\left(\alpha^{(8,1)}_2+ \right. \nonumber\\
&&\hspace{-0.6cm}\left . \frac{2}{(4\pi)^2}\left(1-\frac{K}{N}\right)
\beta^{(8,8)}_3\right)(M_K^2-M_\pi^2) +
\nonumber \\
&&\hspace{-0.6cm}\frac{\alpha^{(8,8)}}{(4\pi)^2}\left(
\sum_{i~valence}M_{3vi}^2(L(M_{3vi})-1) - \right . \nonumber\\
&&\hspace{-0.7cm}\sum_{i~valence}\hspace{-0.3cm}M_{2vi}^2(L(M_{2vi})-1)
-\hspace{-0.2cm}
\sum_{i~sea}\hspace{-0.1cm}M_{3si}^2(L(M_{3si})-1)   \nonumber \\
&&\hspace{-0.6cm}\left.
+\sum_{i~sea}M_{2si}^2(L(M_{2si})-1)\right)\Biggr\}\label{ktovac} \ .
\ea
Here
$L(M)=\log{\frac{M^2}{\Lambda^2}}$, 
and the result is given in the $\overline{MS}$ scheme, with $\Lambda$
the running scale.  $M_{3si}$ ($M_{2si}$) is the mass of a meson made out of
the 3rd (2nd) valence (\ie\ the strange (down)) quark and the $i$th sea quark;
analogously for 
$M_{3vi}$ ($M_{2vi}$) with sea replaced by valence.
At the order we are working, $M_{3si}^2-M_{2si}^2=
M_{3vi}^2-M_{2vi}^2=M_K^2-M_\pi^2$, which follows from
$M_{3si}^2=B_0(m_{v3}+m_{si})$, {\it etc.}
These results also contain the unquenched result as a particular case,  
\ie\ for $N=K=3$ and by equating sea and valence quark masses,
$m_{si}=m_{vi}$, $i=1,\dots,K$.
Quenched ($N=0$) results are obtained by replacing
$\alpha^{(8,1)}_{1,2}\to\alpha^{(8,1)}_{q1,2}$,
$\alpha^{(8,8)}\to\alpha^{NS}_q$ and
$\beta^{(8,8)}_{1,2,3}\to\beta^{NS}_{q1,2,3}$ in 
eqs.~(\ref{ktopi},\ref{ktovac}),
and by dropping all terms containing sea quarks.

We conclude that (in general) the bosonization of the new 
operator $Q^{PQA}_{penguin}$ contributes to a weak matrix element at leading 
order in ChPT with non-analytic terms of the form
$M^2\log{M^2}$, which are absent in the unquenched case, and with analytic 
contributions parameterized by new LECs $\beta^{(8,8)}_{1,2,3}$ ($N\ne 0$)
or $\beta^{NS}_{q1,2,3}$ ($N=0$). 

\section{Lattice strategies}

To leading order in ChPT, the physical $K\to\pi\pi$ 
amplitude is determined by $\alpha_1^{(8,1)}$, and therefore this is the
LEC one wishes to extract from the $K\to\pi$ matrix element.
It is then clear that this can be done by considering only the
singlet penguin, $Q_{penguin}^{PQS}$, in the PQ theory with $N=3$
light sea quarks. This is
equivalent to omitting all Wick contractions in which $q$ and
$\qbar$ in eq.~(\ref{penguin}) are contracted (eye graphs), except when they
correspond to sea quarks.\footnote{The remaining ``eye graphs" are those for 
which $q$ ($\qbar$) is contracted with $\bar{s}$ ($d$) 
in eq.~(\ref{penguin}).}  Of course, since the $K\to\pi$ 
matrix element only gives the linear combination $\alpha_1^{(8,1)}
-\alpha_2^{(8,1)}$, the $K\to 0$ matrix element of 
$Q_{penguin}^{PQS}$ is also needed, as usual \cite{betal}.

In the case that $N\ne 3$, one can still follow a similar strategy
in order to determine $\alpha_1^{(8,1)}$.  However, since the LECs
depend on the number of light sea quarks $N$, there is no reason
why the result should be the same as that of the real world,
which has $N=3$.  
In particular,
in the quenched case different strategies are possible.  If one
assumes that $\alpha_1^{(8,1)}(N=0)\approx\alpha_1^{(8,1)}(N=3)$,
one should use the strategy described above.  Alternatively,
one may include both the singlet and non-singlet operators,
also in the conversion to $K\to\pi\pi$ \cite{gpll}.  Since it is
not known which (if any) of these strategies is a better approximation,
we believe that the difference between them should be interpreted as
an indication of the quenching error.


Finally, we stress that our observations also applies to
$LL$ penguin operators (in which case the new effects start at
order $p^4$ \cite{OUR,gpll}) and, more in general, to
any weak matrix element to which penguin 
operators contribute, such as also non-leptonic $B$ decays.  

\bigskip
We would like to thank the RBC collaboration, Claude Bernard,
Guido Martinelli, Santi Peris and Steve Sharpe for discussions.

\end{document}